\documentclass[
aps, pra,
amsmath,amssymb,
10pt,
final,
tightenlines,
twoside,
twocolumn,
nofloats,
nofootinbib,
superscriptaddress,
showkeys,
showkeywords,
]
{revtex4-2}

\usepackage[T2A]{fontenc}
\usepackage[utf8x]{inputenc}
\usepackage[russian,english]{babel}
\usepackage{graphicx}
\usepackage{dcolumn}
\usepackage{bm}

\input{maik.rty}

\makeatletter
\def\@keys@name{Keywords: }
\makeatother

\setcitestyle{authoryear,round}
\def\squareforqed{\hbox{\rlap{$\sqcap$}$\sqcup$}}

\def\sq{\ifmmode\squareforqed\else{\unskip\nobreak\hfil
\penalty50\hskip1em\null\nobreak\hfil\squareforqed
\parfillskip=0pt\finalhyphendemerits=0\endgraf}\fi}

\def\arcmin{\hbox{$^\prime$}}

\def\utw{\smash{\rlap{\lower5pt\hbox{$\sim$}}}}

\def\udtw{\smash{\rlap{\lower6pt\hbox{$\approx$}}}}

\def\farcs{\hbox{$\,.\!\!^{\prime\prime}$}}

\def\diameter{{\ifmmode\mathchoice
{\ooalign{\hfil\hbox{$\displaystyle/$}\hfil\crcr
{\hbox{$\displaystyle\mathchar"20D$}}}}
{\ooalign{\hfil\hbox{$\textstyle/$}\hfil\crcr
{\hbox{$\textstyle\mathchar"20D$}}}}
{\ooalign{\hfil\hbox{$\scriptstyle/$}\hfil\crcr
{\hbox{$\scriptstyle\mathchar"20D$}}}}
{\ooalign{\hfil\hbox{$\scriptscriptstyle/$}\hfil\crcr
{\hbox{$\scriptscriptstyle\mathchar"20D$}}}}
\else{\ooalign{\hfil/\hfil\crcr\mathhexbox20D}}%
\fi}}

\def\*{$^{*}$}

\newcommand{\MC}{\multicolumn}
\newcommand{\kms}{km\,s$^{-1}$}
\newcommand{\ESO}{ESO\,428$-$05}
\def\fbs{\textsc{fbs}}
\DeclareRobustCommand{\ion}[2]{%
    \relax\ifmmode
    \ifx\testbx\f
    {\mathrm{#1\,\textsc{#2}}}\else
    {\mathrm{#1\,\mathsc{#2}}}\fi
    \else\textup{#1\,{\mdseries\textsc{#2}}}%
    \fi}
\makeatletter
\def\p@subsection{}
\def\p@subsubsection{}
\makeatother

\begin{document}

\selectlanguage{english}


\title{LONG-SLIT SPECTROSCOPY OF THE BIPOLAR PLANETARY NEBULA ESO\,428$-$05}
\author{\firstname{A.~Yu.}~\surname{Kniazev}}
\email{a.kniazev@saao.nrf.ac.za}
\affiliation{South African Astronomical Observatory, Cape Town, 7935, South Africa}
\affiliation{Southern African Large Telescope, Cape Town, 7935, South Africa}
\affiliation{Sternberg Astronomical Institute, Lomonosov Moscow State University, Moscow, 119991, Russia}

\begin{abstract}
The results of detailed long-slit spectroscopy of the bipolar planetary nebula \ESO\ are presented. The observations were obtained with the facility spectrograph RSS of the SALT telescope. The total measured extent of the nebula reaches $\approx$55$^{\prime\prime}$, and the kinematics along the slit directly confirms its bipolar structure. The inclination of the nebula to the line of sight is estimated as $i\approx 45^{\circ}$, and the kinematic ages of two morphological structures are measured: the expanding ring and the polar outflow in the direction perpendicular to the ring. The age of the former is $\approx$1.5$\times$10$^4$~yr and that of the latter is $\approx$(3--4)$\times$10$^4$~yr, i.e. the polar outflow is 2--3 times older than the central ring. The ratio of these ages does not depend on the adopted distance. The weighted mean heliocentric velocity of \ESO\ is V$_{\rm hel}$=74.7$\pm$1.6~\kms. Chemical abundances are obtained for 19 regions along the major axis of the nebula, over an extent of 40$^{\prime\prime}$, i.e. for different morphological parts of \ESO. The abundances were derived with the T$_{\rm e}$--method, where the electron temperature was calculated directly using the measured faint auroral lines [\ion{O}{iii}]~$\lambda$4363~\AA\ and [\ion{N}{ii}]~$\lambda$5755~\AA. The abundances of O, N, Ne, S, Ar, Cl and He were determined. Despite a significant stratification, the measured abundances along the slit can be considered nearly constant.
\end{abstract}

\maketitle

\section*{INTRODUCTION}

Planetary nebulae (PNe) represent the final stage of the evolution of low- and intermediate-mass stars and are one of the main sources of enrichment of the interstellar medium with the products of nuclear burning. Morphologically, PNe are usually divided into round, elliptical, bipolar and point-symmetric ones \citep{1995A&A...293..871C}, and it is the bipolar PNe that occupy a special place among them. They are more strongly concentrated towards the Galactic plane than PNe of the other types, which points to more massive progenitors, and, as a rule, they show enhanced abundances of nitrogen and helium, i.e. they belong to the type~I PNe according to the classification of \citet{1978IAUS...76..215P}. A review of the observational and theoretical aspects of the formation of the various
morphological types of PNe is given in \citet{2002ARA&A..40..439B}.

The reason why a spherically symmetric mass loss on the asymptotic giant branch leads to such asymmetric nebulae remained a matter of debate for a long time. Mechanisms related to stellar rotation and magnetic fields \citep{1999ApJ...517..767G}, as well as to collimated fast outflows and jets \citep{1998AJ....116.1357S}, have been proposed. At present, however, the most convincing one is the hypothesis of the binarity of the central star, when the bipolar structure is formed as a result of the evolution of a binary system through the common-envelope stage \citep[see the reviews by][]{2009PASP..121..316D,2017NatAs...1E.117J,2019ibfe.book.....B}. This hypothesis has also received direct observational support: close binary central stars have been found in a number of bipolar PNe, and the orientation of the nebula turns out to be related to the orbital plane of the binary system \citep{2016ApJ...832..125H}, while the most extreme values of the abundance discrepancy factor are observed precisely in PNe whose central stars are products of common-envelope evolution \citep{2018MNRAS.480.4589W}. A detailed study of the kinematics and chemical composition of individual bipolar PNe
therefore remains an important source of information about these processes.

Several years ago the author started a project of a systematic study of the velocities and chemical abundances of planetary nebulae (PNe) in the direction of the stellar overdensity in the constellation of Canis Major (hereafter CMa), which was interpreted by \citet{2004MNRAS.348...12M} as the remnant of a dwarf galaxy. This study is being carried out with the long-slit Robert Stobie Spectrograph \citep[RSS;][]{2003SPIE.4841.1463B, 2003SPIE.4841.1634K} mounted on the Southern African Large Telescope \citep[hereafter SALT;][]{2006SPIE.6267E..0ZB,2006MNRAS.372..151O}. The first spectroscopic results for PNe in this region were published in \cite{2012AstL...38..707K}. An interesting result obtained in that paper was the finding that the PN \ESO, also known under the names Hen\,2$-$2, PN\,G240.3$-$07.6 and PN\,M\,3$-$2, is, by its properties (metallicity, velocity and position), a likely candidate for membership in the remnants of a possible dwarf galaxy destroyed by tidal interaction with the Milky Way. Since the data obtained in 2006 were not of very high quality and did not cover the full spectral range, the observations of this interesting PN were repeated at a better level, and the results of this study are presented in this paper.

This paper is organized as follows.
Section~\ref{txt:Obs_and_Analysis} describes the observations, the data reduction and the analysis technique.
Section~\ref{txt:results} contains the results of our analysis and their discussion.
The conclusions are presented in Section~\ref{txt:summ}.

\begin{figure}
    \includegraphics[clip=,angle=0,width=\columnwidth]{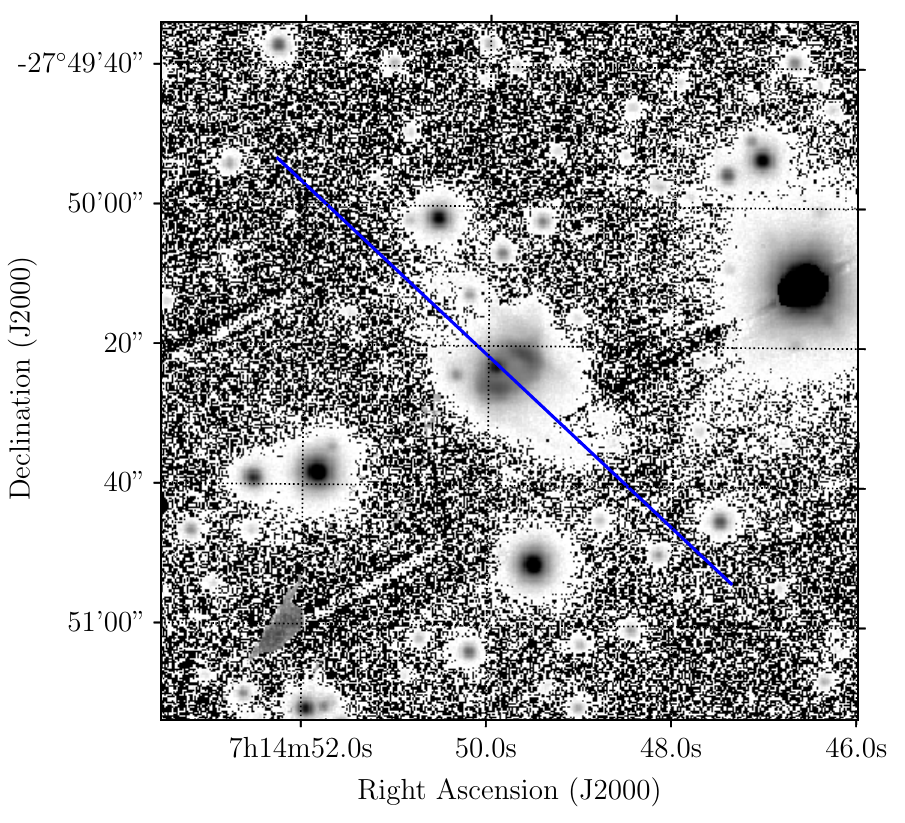}
    \caption{Image of the PN in the r filter taken from the \mbox{Pan-STARRS1} survey
        \citep{2016arXiv161205560C}. The image is fully reduced and is the result of stacking and mosaicking, so that background inhomogeneities and artefacts related to these procedures are visible. Nevertheless, the complex structure of the PN itself is seen, consisting of a central core and additional shells or jets that coincide with the slit position.
        \label{fig:image}}
\end{figure}

\begin{figure*}
    \includegraphics[clip=,angle=0,width=\textwidth]{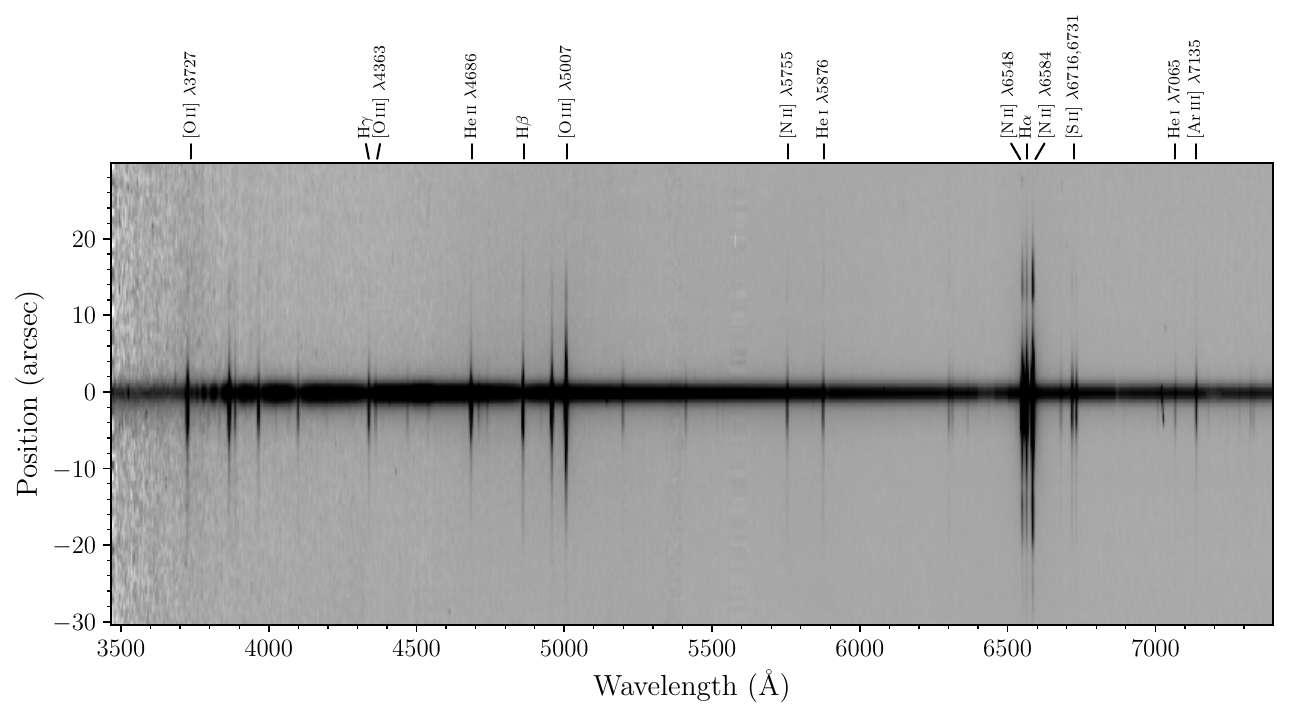}
    \caption{The reduced and combined two-dimensional spectrum. The most important emission lines are indicated at the top. The centre of the slit is tied to the position of the bright star located almost at the centre of \ESO, which is not the central star of this PN (details are given in Section~\ref{txt:central_star}). Positive values of the position along the slit correspond to the North-Eastern direction of the slit in Figure~\ref{fig:image}.
        \label{fig:2D}}
\end{figure*}

\begin{figure*}
    \includegraphics[clip=,angle=0,width=0.85\textwidth]{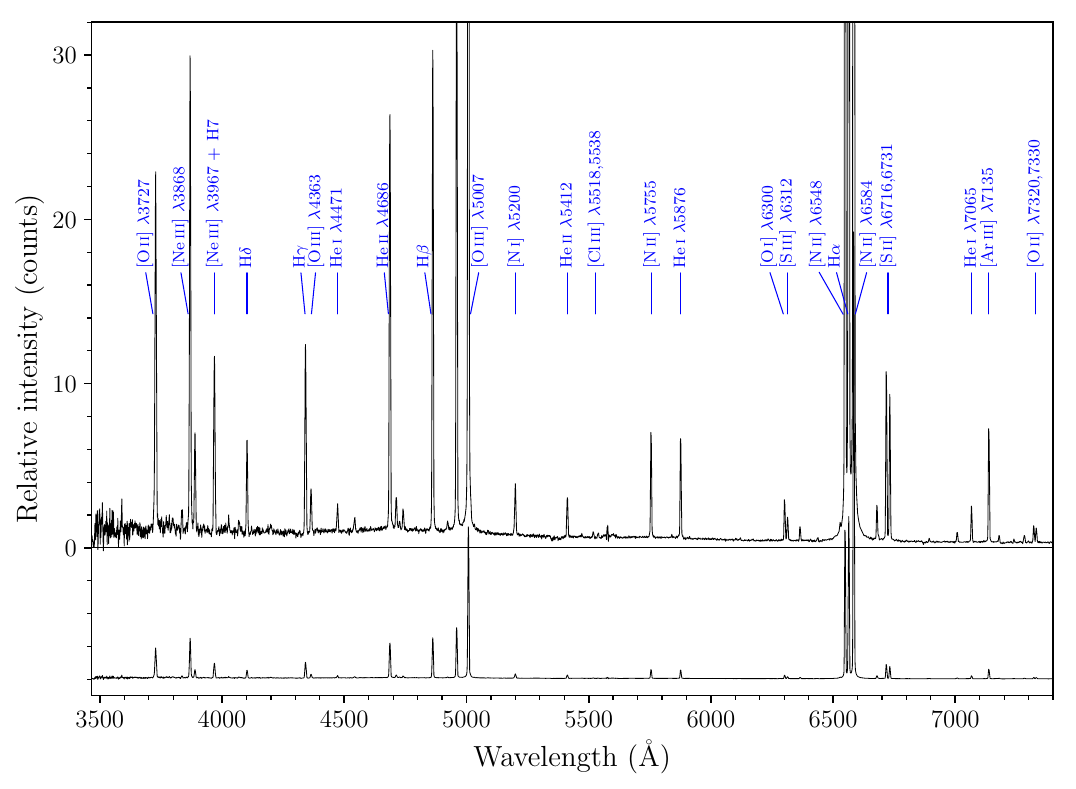}
    \caption{The one-dimensional final spectrum of the region centred at $+3^{\prime\prime}$. All the main emission lines of the nebula are shown. In the lower part of the panel the same spectrum is shown compressed by a factor of 10 and shifted in intensity in order to display the ratios of the strongest lines.
        \label{fig:Region_D}}
\end{figure*}

\section{Spectroscopic observations}
\label{txt:Obs_and_Analysis}

\subsection{Observations and data reduction}
\label{txt:Obs_and_Red}

The observations in the long-slit mode were carried out on 5 January 2013 under a seeing of 1\farcs2. The size of one pixel along the slit is 0\farcs129 and the full slit length is 8\arcmin. During the observations a binning factor of 4 was used along the slit, which resulted in a pixel size along the slit of 0\farcs51. The observations were performed with two spectral setups: the first one covered the range 3470--6600~\AA\ and the second one 4300--7400~\AA. In both setups a short exposure of 60~s was taken first, then a long exposure of 600~s, and afterwards the comparison spectrum and the flat fields. A grating with 900 lines/mm was used, with a reciprocal dispersion of about 0.96~\AA\ per pixel and a spectral resolution $R \approx1100$ (FWHM\,$\sim 4.5$~\AA) for the adopted slit width of 1\farcs25. Xe and Ar arc lamps were used for the wavelength calibration. In order to obtain a correct relative energy distribution, the spectrum of the spectrophotometric standard LTT\,4364 was observed after the observations of the PN, once astronomical twilight had begun. Since SALT is a telescope with a variable pupil, an absolute flux calibration of the spectra is impossible, but the relative spectral energy distribution is correct because the telescope is equipped with an atmospheric dispersion corrector (ADC). During the observations the slit was set at a position angle of 39~degrees, which corresponds to the direction of the major axis of the studied object, as shown in Figure~\ref{fig:image}.

The primary and the subsequent reduction of the spectral data of the RSS spectrograph were performed following the technique described in detail in \cite{2022AstB...77..334K}. The two-dimensional, fully reduced spectra were combined into a common spectrum covering the spectral range 3500--7400~\AA, which made it possible to exclude the gaps between the three CCDs of the mosaic. The final two-dimensional reduced and combined spectrum is shown in Figure~\ref{fig:2D}, where the most important emission lines visible in this two-dimensional spectrum and used in the subsequent analysis are also indicated. In the analysis described in the following sections the two-dimensional spectrum was divided into three-arcsecond apertures with the centre (position 0) at the position of the star that was initially assumed to be the central star of the studied PN. Figure~\ref{fig:Region_D} shows the one-dimensional spectrum of the central part of \ESO\ at the position $+3^{\prime\prime}$ and displays all the main lines of the nebula that were measured and used to calculate the temperature, the electron density and the chemical abundances.

\subsection{Determination of the physical parameters and the chemical composition}
\label{txt:Analysis}

All emission lines were measured using the programs described in detail in \citet{2004ApJS..153..429K}, but rewritten by the author in python \citep{2025RAA....25d5012K}. An important point is the estimation of the total errors of the measured line fluxes as the sum of: (1) the error of the continuum placement, (2) the error related to the Poisson statistics of the photon flux in the line and (3) the error of the sensitivity curve construction (for the observations described here the latter error was about 2\%). All error components were added in quadrature, and the total error of the line intensities was used further by the programs to compute the physical parameters and the chemical composition.

The analysis of the emission-line spectra was carried out within the framework of the classical three-zone model, used by the author many times \citep[for example,][]{2008MNRAS.388.1667K}. With this technique the abundances of the chemical elements O, N, S, Ar, Ne and Cl were determined, together with physical parameters such as the electron temperature ($\rm T_e$) and the electron density ($\rm N_e$). All these parameters and their errors were determined taking into account the errors of the emission-line intensities measured at the previous step. The spectral data obtained for \ESO\ covered the whole required spectral range, and the overlapping spectral setups made it possible to construct a continuous final spectrum free of the empty gaps between the CCD chips of the mosaic. All this allowed us to use two independent electron temperatures in the calculations: T$_e$(\ion{O}{iii}) for the hot zone and T$_e$(\ion{N}{ii}) for the cold zone. The former was calculated from the ratios of the lines [\ion{O}{iii}]~$\lambda$$\lambda$4363,4959,5007~\AA, and the latter from the ratios of the lines [\ion{N}{ii}] $\lambda$$\lambda$5755,6548,6584~\AA.

The heliocentric velocities were calculated following the technique described in detail in \cite{2000A&AS..144..429Z}, in which the measured velocities were corrected for possible instrumental shifts by comparison with known night-sky emission lines measured in the same spectrum.

The spectrum of the central star was analysed using the software package \fbs, developed by the author and collaborators for the study of the spectra of binary stars \citep{2025RAA....25b5024K}. These programs use a library of theoretically computed high-resolution stellar spectra and are designed to determine the radial velocities and the stellar parameters ($T_\mathrm{eff}$, $\log g$, $\mathrm{v \sin i}$, [Fe/H], $E(B-V)$) for each of the components of a binary system. When it is known that the object is a binary star, two model spectra are fitted for the components, each with its own radial velocity and atmospheric parameters, i.e. the observed spectrum is decomposed into the individual spectra of the two components. A single star is a particular case, in which the contribution of the second component is zero. In the present study the theoretical stellar models \textsc{phoenix} \citep{2013A&A...553A...6H} were used, resampled to the resolution of the obtained data, FWHM=4.5~\AA.

\begin{table}
    \caption{Emission line intensities in the region $+3^{\prime\prime}$}
    \label{tab:Lines1}
    {
        \renewcommand{\arraystretch}{1.1}%
        \begin{tabular}{lrr} \hline \hline
            \rule{0pt}{10pt}
            & \MC{2}{c}{Region $+3^{\prime\prime}$}  \\ \hline
            \rule{0pt}{10pt}
            $\lambda_{0}$(\AA) Ion   & $F(\lambda)/F(H\beta)$ & $I(\lambda)/I(H\beta)$  \\ \hline
            3727\ [O\ {\sc ii}]\                          & 144.5$\pm$ 6.8 & 196.7$\pm$ 9.7 \\
            3869\ [Ne\ {\sc iii}]\                        & 139.3$\pm$ 5.9 & 181.5$\pm$ 8.0 \\
            3889\ He\ {\sc i}\ +\ H8\                     &  27.9$\pm$ 2.5 &  36.2$\pm$ 3.3 \\
            3967\ [Ne\ {\sc iii}]\ +\ H7\                 &  44.6$\pm$ 2.6 &  56.5$\pm$ 3.3 \\
            4026\ He\ {\sc i}\                            &   3.7$\pm$ 1.1 &   4.6$\pm$ 1.3 \\
            4069\ [S\ {\sc ii}]\                          &   8.8$\pm$ 1.4 &  10.8$\pm$ 1.7 \\
            4102\ H$\delta$\                              &  22.3$\pm$ 1.5 &  27.2$\pm$ 1.9 \\
            4340\ H$\gamma$\                              &  39.3$\pm$ 1.9 &  44.8$\pm$ 2.2 \\
            4363\ [O\ {\sc iii}]\                         &  10.0$\pm$ 1.1 &  11.3$\pm$ 1.2 \\
            4471\ He\ {\sc i}\                            &   5.4$\pm$ 0.8 &   5.9$\pm$ 0.9 \\
            4686\ He\ {\sc ii}\                           &  75.7$\pm$ 2.8 &  79.0$\pm$ 2.9 \\
            4712\ [Ar\ {\sc iv}]\ +\ He\ {\sc i}\         &   6.4$\pm$ 0.9 &   6.6$\pm$ 0.9 \\
            4740\ [Ar\ {\sc iv}]\                         &   7.3$\pm$ 0.8 &   7.5$\pm$ 0.9 \\
            4861\ H$\beta$\                               & 100.0$\pm$ 1.6 & 100.0$\pm$ 1.7 \\
            4959\ [O\ {\sc iii}]\                         & 136.6$\pm$ 4.4 & 133.5$\pm$ 4.3 \\
            5007\ [O\ {\sc iii}]\                         & 409.3$\pm$11.2 & 395.7$\pm$10.9 \\
            5200\ [N\ {\sc i}]\                           &  15.2$\pm$ 0.9 &  14.1$\pm$ 0.9 \\
            5518\ [Cl\ {\sc iii}]\                        &   0.9$\pm$ 0.3 &   0.8$\pm$ 0.3 \\
            5538\ [Cl\ {\sc iii}]\                        &   0.4$\pm$ 0.2 &   0.3$\pm$ 0.2 \\
            5755\ [N\ {\sc ii}]\                          &  25.7$\pm$ 1.1 &  21.2$\pm$ 0.9 \\
            5876\ He\ {\sc i}\                            &  26.3$\pm$ 1.1 &  21.2$\pm$ 0.9 \\
            6300\ [O\ {\sc i}]\                           &   9.8$\pm$ 0.6 &   7.3$\pm$ 0.4 \\
            6312\ [S\ {\sc iii}]\                         &   5.6$\pm$ 0.4 &   4.1$\pm$ 0.3 \\
            6364\ [O\ {\sc i}]\                           &   3.3$\pm$ 0.3 &   2.4$\pm$ 0.3 \\
            6548\ [N\ {\sc ii}]\                          & 389.7$\pm$11.3 & 280.1$\pm$ 8.9 \\
            6563\ H$\alpha$\                              & 385.7$\pm$11.4 & 276.6$\pm$ 8.9 \\
            6583\ [N\ {\sc ii}]\                          &1121.1$\pm$28.8 & 801.3$\pm$23.1 \\
            6678\ He\ {\sc i}\                            &   7.4$\pm$ 0.5 &   5.2$\pm$ 0.4 \\
            6716\ [S\ {\sc ii}]\                          &  39.2$\pm$ 0.7 &  27.4$\pm$ 0.6 \\
            6731\ [S\ {\sc ii}]\                          &  32.7$\pm$ 0.6 &  22.8$\pm$ 0.5 \\
            6891\ He\ {\sc ii}\                           &   0.4$\pm$ 0.2 &   0.2$\pm$ 0.1 \\
            7065\ He\ {\sc i}\                            &   6.5$\pm$ 0.4 &   4.3$\pm$ 0.3 \\
            7136\ [Ar\ {\sc iii}]\                        &  22.1$\pm$ 0.9 &  14.6$\pm$ 0.6 \\
            7178\ He\ {\sc ii}\                           &   1.1$\pm$ 0.2 &   0.7$\pm$ 0.1 \\
            7281\ He\ {\sc i}\                            &   0.9$\pm$ 0.2 &   0.6$\pm$ 0.2 \\
            7320\ [O\ {\sc ii}]\                          &   3.0$\pm$ 0.3 &   2.0$\pm$ 0.2 \\
            7330\ [O\ {\sc ii}]\                          &   1.8$\pm$ 0.2 &   1.1$\pm$ 0.1 \\
            && \\
            C(H$\beta$)\ dex                              & \MC {2}{c}{0.43$\pm$0.04}             \\
            E(B-V)\ mag                                   & \MC {2}{c}{0.30$\pm$0.03}             \\
            \hline\hline
        \end{tabular}
    }
\end{table}

\begin{table}
    \centering{
        \caption{Physical conditions and abundances in the region $+3^{\prime\prime}$}
        \label{t:Chem1}
        \renewcommand{\arraystretch}{1.1}%
        \begin{tabular}{lc} \hline\hline
            \rule{0pt}{10pt}
            Quantity                             &  Region $+3^{\prime\prime}$ \\
            $T_{\rm e}$(OIII)(K)\                &  18185$\pm$1092 ~~     \\
            $T_{\rm e}$(OII)(K)\                 &  14904$\pm$131 ~~      \\
            $T_{\rm e}$(NII)(K)\                 &  13433$\pm$778 ~~      \\
            $T_{\rm e}$(SIII)(K)\                &  17283$\pm$720 ~~      \\
            $N_{\rm e}$(SII)(cm$^{-3}$)\         &  253$\pm$82~~          \\
                                                 &                        \\
            O$^{+}$/H$^{+}$($\times$10$^5$)\     &  1.613$\pm$0.307~~     \\
            O$^{++}$/H$^{+}$($\times$10$^5$)\    &  2.775$\pm$0.371~~     \\
            O$^{+++}$/H$^{+}$($\times$10$^5$)\   &  2.021$\pm$0.272~~     \\
            O/H($\times$10$^5$)\                 &  6.409$\pm$0.553~~     \\
            12+log(O/H)\                         &  ~7.81$\pm$0.04~~      \\
                                                 &                        \\
            N$^{+}$/H$^{+}$($\times$10$^7$)\     &  769.50$\pm$96.58~~    \\
            ICF(N)\                              &  3.893                 \\
            N/H($\times$10$^5$)\                 &  29.96$\pm$3.76~~      \\
            12+log(N/H)\                         &  8.48$\pm$0.05~~       \\
            log(N/O)\                            &  0.67$\pm$0.07~~       \\
                                                 &                        \\
            Ne$^{++}$/H$^{+}$($\times$10$^5$)\   &  2.889$\pm$0.450~~     \\
            ICF(Ne)\                             &  1.255                 \\
            Ne/H($\times$10$^5$)\                &  3.626$\pm$0.565~~     \\
            12+log(Ne/H)\                        &  7.56$\pm$0.07~~       \\
            log(Ne/O)\                           &  $-$0.25$\pm$0.08~~    \\
                                                 &                        \\
            S$^{+}$/H$^{+}$($\times$10$^7$)\     &  1.587$\pm$0.318~~     \\
            S$^{++}$/H$^{+}$($\times$10$^7$)\    &  14.17$\pm$1.93~~      \\
            ICF(S)\                              &  1.134                 \\
            S/H($\times$10$^7$)\                 &  17.87$\pm$2.22~~      \\
            12+log(S/H)\                         &  6.25$\pm$0.05~~       \\
            log(S/O)\                            &  $-$1.55$\pm$0.07~~    \\
                                                 &                        \\
            Ar$^{++}$/H$^{+}$($\times$10$^7$)\   &  4.623$\pm$0.362~~     \\
            Ar$^{+++}$/H$^{+}$($\times$10$^7$)\  &  4.767$\pm$0.554~~     \\
            ICF(Ar)\                             &  1.010                 \\
            Ar/H($\times$10$^7$)\                &  9.485$\pm$0.669~~     \\
            12+log(Ar/H)\                        &  5.98$\pm$0.03~~       \\
            log(Ar/O)\                           &  $-$1.83$\pm$0.05~~    \\
                                                 &                        \\
            Cl$^{++}$/H$^{+}$($\times$10$^7$)\   &  0.178$\pm$0.053~~     \\
            ICF(Cl)\                             &  1.345                 \\
            Cl/H($\times$10$^7$)\                &  0.240$\pm$0.072~~     \\
            12+log(Cl/H)\                        &  4.38$\pm$0.13~~       \\
            log(Cl/O)\                           &  $-$3.43$\pm$0.14~~    \\
                                                 &                        \\
            He/H\                                &  0.229$\pm$0.007~~     \\
            \hline\hline
        \end{tabular}
    }
\end{table}

\begin{figure*}
    \includegraphics[clip=,angle=0,width=0.75\textwidth]{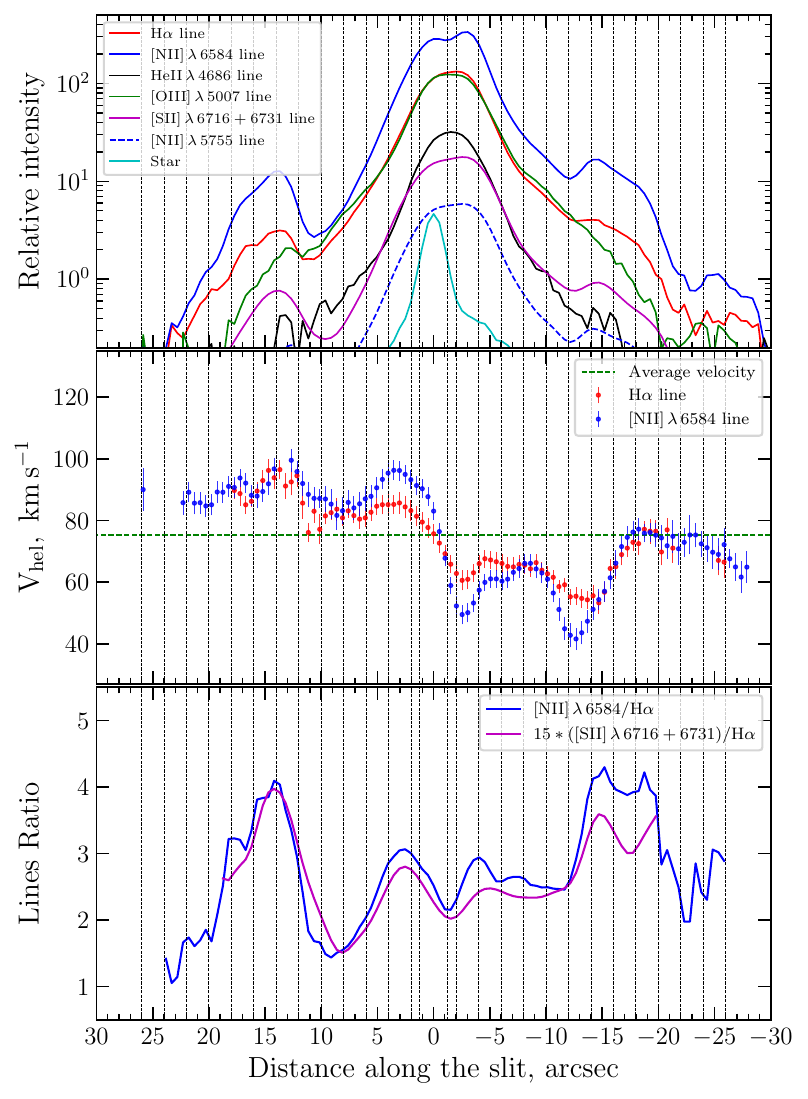}
    \caption{Distribution of the intensities of the brightest emission lines along the slit (upper panel), distribution along the slit of the measured heliocentric velocities (middle panel) and the line ratios [\ion{N}{ii}]~$\lambda$6584/H$\alpha$ and 15*[\ion{S}{ii}]~$\lambda$6716+6731/H$\alpha$ (lower panel). The upper panel also shows the profile of the bright star to which the position 0 along the slit is tied. In all panels the vertical lines show the boundaries of the apertures within which the spectrum was summed for the subsequent analysis of the chemical abundances and the physical parameters along the slit. Positive values of the position along the slit correspond to the North-Eastern direction of the slit in Figure~\ref{fig:image}.
        \label{fig:vel}}
\end{figure*}
\begin{figure*}
    \includegraphics[clip=,angle=0,width=0.75\textwidth]{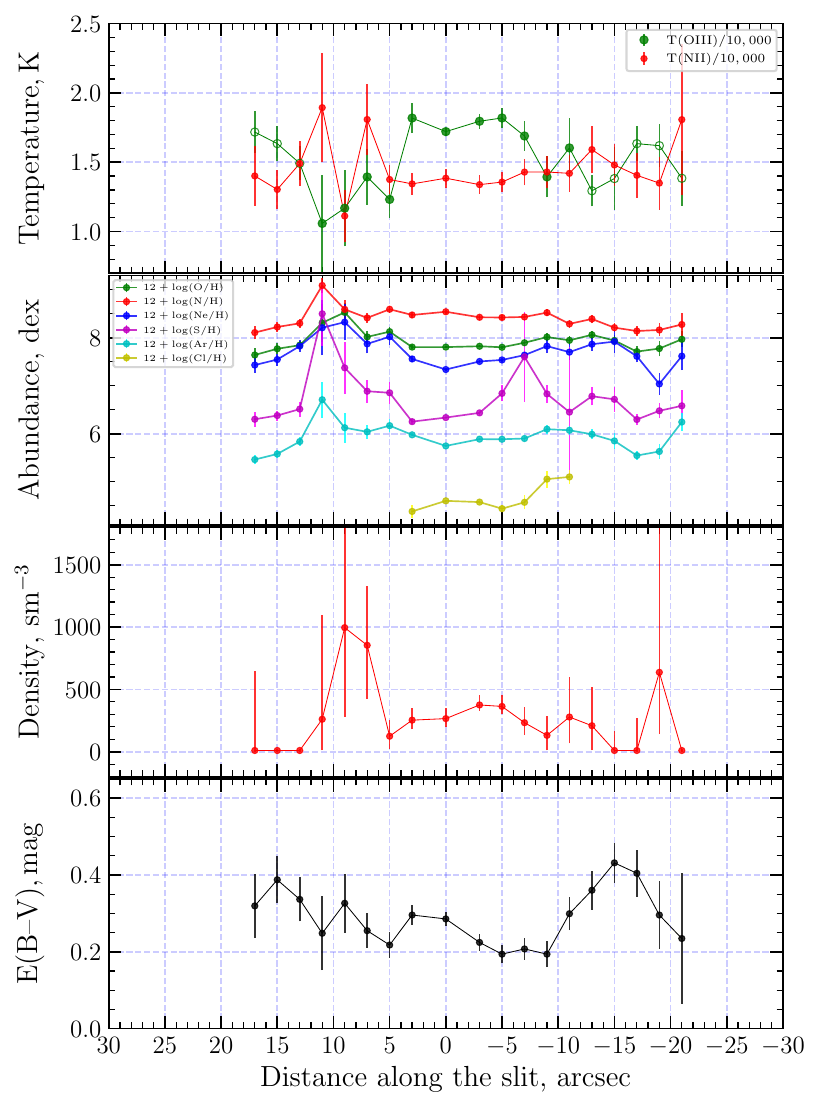}
    \caption{Distribution of the derived physical parameters and chemical abundances along the slit. Positive values of the position along the slit correspond to the North-Eastern direction of the slit in Figure~\ref{fig:image}.
        \label{fig:2D_par}}
\end{figure*}

\section{Results}
\label{txt:results}

An example of the measured relative intensities of the emission lines for the spectrum of the region centred at $+3^{\prime\prime}$, shown in Figure~\ref{fig:Region_D}, together with the values of the extinction coefficient C(H$\beta$) and the derived values of the colour excess $E(B-V)$, is given in Table~\ref{tab:Lines1}. The derived electron temperatures T$_{\rm e}$, the electron density N$_{\rm e}$, and the derived abundances of oxygen, nitrogen, neon, sulphur, argon, chlorine and helium for this region are given in Table~\ref{t:Chem1}. Table~\ref{tab:Major_par} contains the set of the derived physical parameters of the PN along the slit, and Table~\ref{tab:Major_che} contains the set of the chemical abundances along the slit (along the major axis) of the studied PN.
The distribution along the slit of the intensities of the various emission lines, of the velocities and of the line ratios [\ion{N}{ii}]~$\lambda$6584/H$\alpha$ and [\ion{S}{ii}]~$\lambda$6716+6731/H$\alpha$ is shown in Figure~\ref{fig:vel}.
The distribution of these abundances and parameters along the slit is shown in Figure~\ref{fig:2D_par}.

\begin{table*}
    \centering{
        \caption{Physical parameters of \ESO\ along the slit. The position is measured from
            the bright star projected close to the centre of the nebula; positive
            values correspond to the North-Eastern direction}
        \label{tab:Major_par}
        \renewcommand{\arraystretch}{1.1}%
        \begin{tabular}{rccccc} \hline\hline
            \rule{0pt}{10pt}
            Position                     & C(H$\beta$)     & $E(B-V)$        &
            $T_{\rm e}$(\ion{O}{iii})    & $T_{\rm e}$(\ion{N}{ii})          &
            $N_{\rm e}$(\ion{S}{ii})     \\
            (arcsec)                     & (dex)           & (mag)           &
            (K)                          & (K)             & (cm$^{-3}$)     \\ \hline
            $-21$ & 0.34$\pm$0.25 & 0.23$\pm$0.17 & 13839$\pm$1976$^{a}$ & 18072$\pm$5426  & $<$100 \\
            $-19$ & 0.43$\pm$0.13 & 0.30$\pm$0.09 & 16192$\pm$1544$^{a}$ & 13487$\pm$1922  & 637$\pm$906 \\
            $-17$ & 0.59$\pm$0.09 & 0.40$\pm$0.06 & 16338$\pm$1273$^{a}$ & 14053$\pm$1622  & $<$100 \\
            $-15$ & 0.64$\pm$0.08 & 0.43$\pm$0.05 & 13817$\pm$2273$^{a}$ & 14797$\pm$1510  & $<$100 \\
            $-13$ & 0.53$\pm$0.07 & 0.36$\pm$0.05 & 12935$\pm$1130$^{a}$ & 15913$\pm$1713  & 209$\pm$254 \\
            $-11$ & 0.44$\pm$0.06 & 0.30$\pm$0.04 & 16029$\pm$2189       & 14195$\pm$1346  & 278$\pm$267 \\
            $-9$  & 0.28$\pm$0.05 & 0.19$\pm$0.03 & 13936$\pm$1479       & 14291$\pm$1122  & 131$\pm$144 \\
            $-7$  & 0.30$\pm$0.04 & 0.21$\pm$0.03 & 16890$\pm$1062       & 14285$\pm$952   & 233$\pm$110 \\
            $-5$  & 0.28$\pm$0.04 & 0.19$\pm$0.02 & 18192$\pm$706        & 13564$\pm$729   & 363$\pm$75 \\
            $-3$  & 0.33$\pm$0.03 & 0.22$\pm$0.02 & 17955$\pm$556        & 13379$\pm$675   & 375$\pm$60 \\
            ~~0   & 0.42$\pm$0.03 & 0.29$\pm$0.02 & 17206$\pm$285        & 13843$\pm$697   & 266$\pm$72 \\
            $+3$  & 0.43$\pm$0.04 & 0.30$\pm$0.03 & 18185$\pm$1092       & 13433$\pm$778   & 253$\pm$82 \\
            $+5$  & 0.32$\pm$0.05 & 0.22$\pm$0.03 & 12311$\pm$1328       & 13749$\pm$1017  & 124$\pm$120 \\
            $+7$  & 0.38$\pm$0.07 & 0.26$\pm$0.04 & 13935$\pm$2013       & 18081$\pm$2543  & 854$\pm$516 \\
            $+9$  & 0.48$\pm$0.11 & 0.33$\pm$0.08 & 11679$\pm$2765       & 11115$\pm$1902  & 995$\pm$2054 \\
            $+11$ & 0.36$\pm$0.14 & 0.25$\pm$0.10 & 10578$\pm$3481       & 18937$\pm$3924  & 261$\pm$665 \\
            $+13$ & 0.49$\pm$0.08 & 0.34$\pm$0.06 & 14919$\pm$1214$^{a}$ & 14901$\pm$1638  & $<$100 \\
            $+15$ & 0.57$\pm$0.09 & 0.39$\pm$0.06 & 16341$\pm$1281$^{a}$ & 13031$\pm$1435  & $<$100 \\
            $+17$ & 0.47$\pm$0.12 & 0.32$\pm$0.08 & 17174$\pm$1514$^{a}$ & 14007$\pm$2136  & $<$100 \\
            \hline\hline
            \MC{6}{l}{\rule{0pt}{12pt}$^{a}$ The auroral line [\ion{O}{iii}]~$\lambda$4363~\AA\ is not detected;} \\
            \MC{6}{l}{~~~$T_{\rm e}$(\ion{O}{iii}) is not a direct measurement.} \\
        \end{tabular}
    }
\end{table*}
\begin{table*}
    \centering{
        \caption{Chemical abundances in \ESO\ along the slit.
            The position is measured in the same way as in Table~\ref{tab:Major_par}}
        \label{tab:Major_che}
        \renewcommand{\arraystretch}{1.1}%
        \begin{tabular}{rcccccc} \hline\hline
            \rule{0pt}{10pt}
            Position    & 12+log(O/H) & 12+log(N/H) & 12+log(Ne/H) &
            12+log(S/H) & 12+log(Ar/H) & 12+log(Cl/H) \\
            (arcsec)    & (dex)       & (dex)       & (dex)        &
            (dex)       & (dex)        & (dex)        \\ \hline
            $-21$ & 7.97$\pm$0.17 & 8.28$\pm$0.24 & 7.62$\pm$0.29 & 6.58$\pm$0.34 & 6.24$\pm$0.18 & \MC{1}{c}{---} \\
            $-19$ & 7.78$\pm$0.16 & 8.16$\pm$0.14 & 7.04$\pm$0.23 & 6.48$\pm$0.16 & 5.63$\pm$0.16 & \MC{1}{c}{---} \\
            $-17$ & 7.71$\pm$0.11 & 8.14$\pm$0.11 & 7.61$\pm$0.13 & 6.30$\pm$0.12 & 5.55$\pm$0.09 & \MC{1}{c}{---} \\
            $-15$ & 7.94$\pm$0.11 & 8.21$\pm$0.09 & 7.92$\pm$0.23 & 6.72$\pm$0.26 & 5.85$\pm$0.17 & \MC{1}{c}{---} \\
            $-13$ & 8.06$\pm$0.08 & 8.39$\pm$0.09 & 7.87$\pm$0.14 & 6.78$\pm$0.19 & 5.99$\pm$0.11 & \MC{1}{c}{---} \\
            $-11$ & 7.95$\pm$0.09 & 8.29$\pm$0.09 & 7.70$\pm$0.17 & 6.45$\pm$1.35 & 6.07$\pm$0.09 & 5.10$\pm$0.15 \\
            $-9$  & 8.02$\pm$0.08 & 8.53$\pm$0.07 & 7.83$\pm$0.15 & 6.83$\pm$0.18 & 6.09$\pm$0.09 & 5.05$\pm$0.18 \\
            $-7$  & 7.90$\pm$0.04 & 8.44$\pm$0.06 & 7.64$\pm$0.08 & 7.60$\pm$0.94 & 5.90$\pm$0.05 & 4.57$\pm$0.15 \\
            $-5$  & 7.80$\pm$0.03 & 8.42$\pm$0.05 & 7.54$\pm$0.04 & 6.84$\pm$0.17 & 5.89$\pm$0.02 & 4.44$\pm$0.08 \\
            $-3$  & 7.82$\pm$0.02 & 8.43$\pm$0.05 & 7.51$\pm$0.04 & 6.44$\pm$0.05 & 5.89$\pm$0.02 & 4.57$\pm$0.04 \\
            ~~0   & 7.81$\pm$0.02 & 8.54$\pm$0.05 & 7.34$\pm$0.02 & 6.34$\pm$0.02 & 5.75$\pm$0.02 & 4.60$\pm$0.02 \\
            $+3$  & 7.81$\pm$0.04 & 8.48$\pm$0.05 & 7.56$\pm$0.07 & 6.25$\pm$0.05 & 5.98$\pm$0.03 & 4.38$\pm$0.13 \\
            $+5$  & 8.13$\pm$0.09 & 8.60$\pm$0.07 & 8.02$\pm$0.16 & 6.85$\pm$0.23 & 6.17$\pm$0.14 & \MC{1}{c}{---} \\
            $+7$  & 8.02$\pm$0.12 & 8.42$\pm$0.11 & 7.87$\pm$0.20 & 6.89$\pm$0.24 & 6.04$\pm$0.14 & \MC{1}{c}{---} \\
            $+9$  & 8.53$\pm$0.19 & 8.59$\pm$0.19 & 8.33$\pm$0.38 & 7.37$\pm$0.54 & 6.13$\pm$0.31 & \MC{1}{c}{---} \\
            $+11$ & 8.31$\pm$0.39 & 9.09$\pm$0.16 & 8.21$\pm$0.57 & 8.50$\pm$0.67 & 6.71$\pm$0.37 & \MC{1}{c}{---} \\
            $+13$ & 7.85$\pm$0.08 & 8.30$\pm$0.10 & 7.83$\pm$0.12 & 6.51$\pm$0.16 & 5.84$\pm$0.09 & \MC{1}{c}{---} \\
            $+15$ & 7.77$\pm$0.11 & 8.23$\pm$0.11 & 7.55$\pm$0.12 & 6.38$\pm$0.10 & 5.58$\pm$0.09 & \MC{1}{c}{---} \\
            $+17$ & 7.64$\pm$0.15 & 8.11$\pm$0.14 & 7.43$\pm$0.16 & 6.30$\pm$0.16 & 5.46$\pm$0.10 & \MC{1}{c}{---} \\ \hline
            \rule{0pt}{11pt}
            Mean$^{a}$ & 7.84$\pm$0.02 & 8.43$\pm$0.03 & 7.45$\pm$0.04 & 6.37$\pm$0.03 & 5.86$\pm$0.03 & 4.59$\pm$0.03 \\
            $\chi^2/\nu$ (X/H)$^{b}$ & 3.0 & 3.4 & 6.0 & 3.2 & 6.8 & 4.1 \\
            $\chi^2/\nu$ (X/O)$^{c}$ & \MC{1}{c}{---} & 1.9 & 2.3 & 1.4 & 1.9 & 2.0 \\
            \hline\hline
            \MC{7}{l}{\rule{0pt}{12pt}$^{a}$ Weighted mean over all apertures of the logarithmic values 12+log($X$/H) with weights $1/\sigma^2$.} \\
            \MC{7}{l}{~~~The error of the mean is multiplied by $\sqrt{\chi^2/\nu}$ in order to account for the observed scatter.} \\
            \MC{7}{l}{$^{b}$ Reduced $\chi^2$ for the hypothesis of a constant abundance along the slit, computed} \\
            \MC{7}{l}{~~~with respect to the mean given above; $\nu = N-1$ (see Section~\ref{txt:struc_par}).} \\
            \MC{7}{l}{$^{c}$ The same for the ratios log(X/O), in which the shift common to all elements,} \\
            \MC{7}{l}{~~~caused by the error of $T_{\rm e}$, largely cancels out.} \\
        \end{tabular}
    }
\end{table*}

According to its classification, the planetary nebula \ESO\ belongs to the bipolar nebulae of type~I \citep{1978IAUS...76..215P,1996A&AS..116...95L}. Its first spectrum was obtained with the 1.52-m ESO telescope equipped with a photon counter \cite{1991A&AS...90...89A} and covered the spectral region 4000--7400~\AA. The spectral line [\ion{O}{iii}]~$\lambda$4363~\AA\ was not detected in that work, and therefore the temperature $T_{\rm e}$ was not determined. The oxygen abundance was determined as O/H=8.22~dex, while the estimates of the He, N, and S abundances were flagged as extremely uncertain.
In 1994, also with the 1.52-m ESO telescope but already with a CCD detector, a spectrum of the central part of the nebula was obtained and published by \cite{1998A&A...332..721P}, but the emission-line intensities differed significantly from the subsequent measurements. The oxygen abundance was determined as O/H=$7.95\pm0.19$~dex. \cite{2010ApJ...711..619M} published data obtained
in 2003 with the 1.5-m CTIO telescope (Chile) in the spectral region 3600--9600~\AA, while the most recent spectral data were published in \cite{2012AstL...38..707K,2018A&A...619A..84B}. The results of these last papers agree within $\pm1.5\sigma$ of the errors for most of the elements.

In the present work the physical conditions and the abundances of the various chemical elements have for the first time been computed for different regions of the spectrum of the PN \ESO\ along its major axis, and hence for its different morphological parts: the central region, the ring and the regions of the bipolar outflow, which will be discussed in detail in Section~\ref{txt:structure}.

The extent of \ESO\ was estimated in \citet{1992secg.book.....A} as 7.6$^{\prime\prime}$, while in \cite{2012AstL...38..707K} its total extent along the spectrum was estimated as 24$^{\prime\prime}$. The distribution of the emission-line intensity along the slit for the data of the present work can be seen in Figure~\ref{fig:2D} as well as in Figure~\ref{fig:vel}. Both figures show that the central part of \ESO\ indeed has a size of about 8$^{\prime\prime}$; however, the total extent of the nebula in the bright lines of nitrogen [\ion{N}{ii}]~$\lambda$6584~\AA\ and hydrogen H$\alpha$ is close to one arcminute ($\approx$55$^{\prime\prime}$).

An estimate of the measured heliocentric velocity V$_{\rm hel}$=67.7$\pm$15.0~\kms\ was published in \citet{1998A&AS..132...13D} and agrees within the errors with the velocity estimate V$_{\rm hel}$=68.3$\pm$8.2~\kms\ of \citet{2012AstL...38..707K}. The middle panel of Figure~\ref{fig:vel} shows the distribution of the velocity of \ESO\ along the slit, measured from the brightest lines H$\alpha$ and [\ion{N}{ii}]~$\lambda$6584~\AA. These data directly show that \ESO\ has a bipolar structure and that the velocity estimate depends on which part of the nebula fell into the slit of the spectrograph. The weighted mean velocity of \ESO\ from our data is V$_{\rm hel}$=74.7$\pm$1.6~\kms, and this value is shown as a horizontal solid line in the middle panel of Figure~\ref{fig:vel}. This velocity corresponds to the position 0$^{\prime\prime}$ in all panels of that figure, as well as to the position of the ``central star'' (CS) of \ESO\ (this is described in more detail in Section~\ref{txt:central_star}).

\begin{figure*}
    \includegraphics[clip=,angle=0,width=0.9\textwidth]{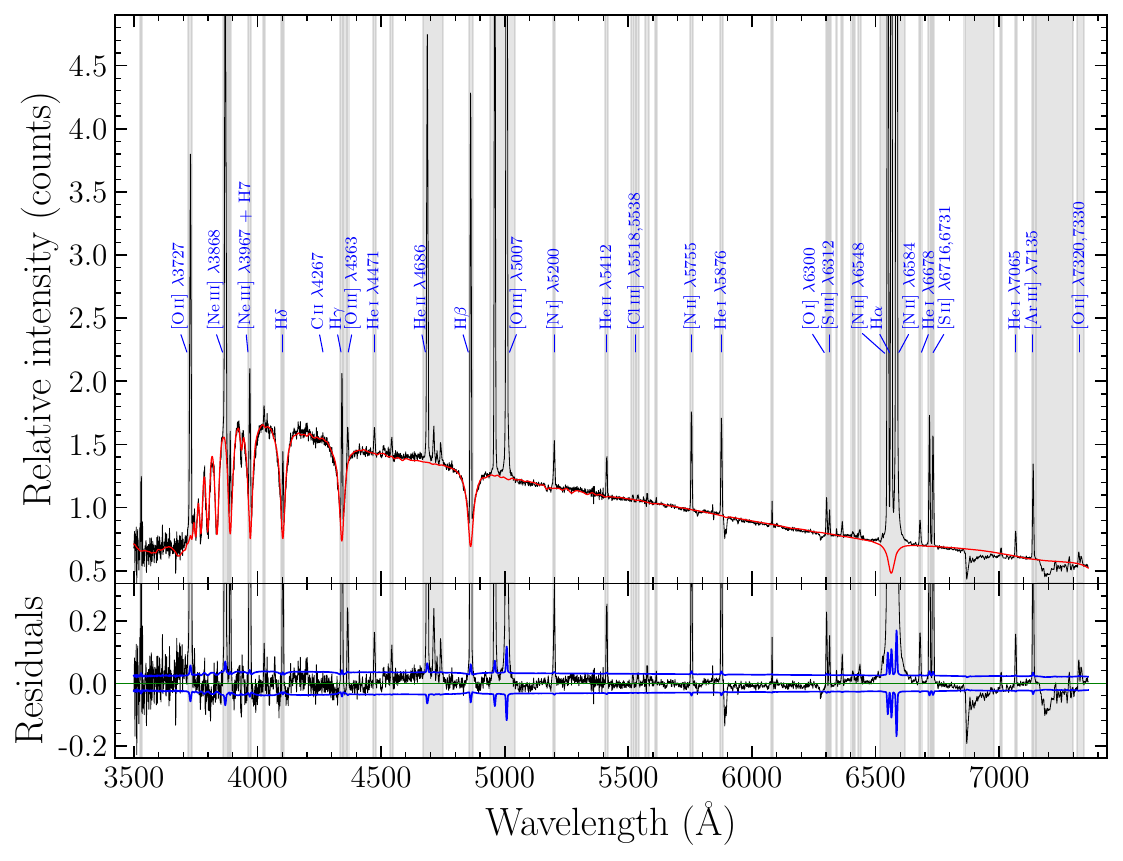}
    \caption{The one-dimensional final spectrum of the central region 0$^{\prime\prime}$, which also contains the spectrum of the possible central star. All the main emission lines of the nebula are shown, the measurements of which are given in Table~\ref{tab:Lines1}. In the upper panel of the figure the model spectrum of the central star with the parameters listed in Table~\ref{tab:Star} is also shown in red. The lower panel shows the difference between the total spectrum and the model spectrum of the star. The 1$\sigma$ total error for each point of the spectrum, which was used in the construction of the model, is also given. In both panels the grey vertical bands show the regions of the observed spectrum that were excluded from the modelling because of the presence of nebular emission lines in them.
        \label{fig:Region_E}}
\end{figure*}

\begin{table}
    \centering{
        \caption{Parameters of the central star of \ESO\ derived from the model of its
            spectrum}
        \label{tab:Star}
        \renewcommand{\arraystretch}{1.1}%
        \begin{tabular}{lr} \hline\hline
            Parameter            & \MC{1}{c}{Value}  \\ \hline
            T$_{\rm eff}$ [K]    &   8100$\pm$200    \\
            log g         [dex ] &   4.40$\pm$0.20   \\
            $\rm [Fe/H]$  [dex]  &$-$0.78$\pm$0.30   \\
            V$_{\rm hel}$ [km/s] &     41$\pm$8      \\
            $E(B-V)$  [mag]      &   0.10$\pm$0.03   \\ \hline\hline
        \end{tabular}
    }
\end{table}

\section{Discussion}
\label{txt:disc}

\subsection{The central star of \ESO}
\label{txt:central_star}

Is the star located close to the centre of \ESO\ the central star of this PN? Or is this star merely projected sufficiently close to the centre while in fact it is not the central star and should not be included in the analysis? The answer to this question has already been given in \citet{2018A&A...619A..84B}, which is specifically devoted to this issue and where it is shown that the star is a foreground binary star with both components having similar effective temperatures T$_{\rm eff}$=\,8100--8500~K and log\,g=\,4.20--4.6~dex. Nevertheless, since the slit of the spectrograph passed through the star, our spectrum allows us to perform an independent analysis as a test of our technique.

The spectrum of the star and its model, obtained with the method described in Section~\ref{txt:Analysis}, are shown in Figure~\ref{fig:Region_E}. Since the package \fbs\ allows any number of spectral intervals of any size to be excluded from the modelling, practically all the parts of the spectrum in which clear nebular emission lines were present, as well as the regions of telluric atmospheric lines, were excluded from the analysis. The extinction $E(B-V)$ is also one of the fitted parameters. A large spectral range helps to determine this parameter more accurately, while the accuracy of the correction for the spectral sensitivity places tight constraints on the result obtained.

The lower panel of Figure~\ref{fig:Region_E} shows that the difference between the model and the observed spectrum is smaller than or equal to the $1\sigma$ total error over the whole spectral range used, and that the model also describes the region of the Balmer jump very well, which attests to the quality of the spectral data.

The parameters of the star, obtained as a result of the analysis with the program \fbs, are given in Table~\ref{tab:Star}. As follows from the table, the parameters T$_{\rm eff}$ and log\,g of the star obtained from the analysis of our data agree with the results of \citet{2018A&A...619A..84B}. Since the spectral resolution is low, we do not resolve the binarity of the star, and we therefore assume that the spectrum contains two spectra of practically identical components. In addition, our analysis shows that the extinction $E(B-V)$ and the heliocentric velocity V$_{\rm hel}$ of the star differ very strongly from the parameters obtained using the spectrum of the PN region 0$^{\prime\prime}$, which was obtained as the difference between the observed spectrum and the model spectrum of the star. These facts, together with the non-central position of the star, indicate unambiguously and independently that this star is a nearby foreground star.

It is worth noting that even the latest Gaia DR3 data release \citep{2016A&A...595A...1G,2020arXiv201201533G} does not allow a sufficiently accurate estimate of the distance to this foreground star, since the absolute stellar parallax of the star from these data is 0.0771$\pm$0.0411~milliarcseconds, which corresponds to a distance to the star of 13$\pm$7~kpc. However, \citet{2018A&A...619A..84B}, from the results of photometric studies, determined the distance to this binary system as 7.5$\pm$0.6~kpc, which on the whole coincides with the distance to the stellar overdensity in the direction of the constellation of Canis Major ($\sim$7.2~kpc). By itself this coincidence, of course, does not establish the membership of the star in CMa, since a significant number of disc stars are projected along the line of sight in this direction, but it does show that this binary system is located substantially closer than the studied PN and is in no way related to it.

\subsection{The distance to \ESO}
\label{txt:distance}

Since the star discussed in the previous section is not the central star of \ESO, the distance to it is not an estimate of the distance to this PN. This circumstance has direct consequences for the literature estimates of the distance. Thus, in the catalogue of \citet{2020ApJ...889...21S}, based on Gaia~DR2 parallaxes for the central stars of PNe, the source Gaia~DR2\,5609860130542365824 is assigned to the object PN\,G240.3$-$07.6. Its coordinates and stellar magnitude ($G = 16.26^{\rm m}$) indicate unambiguously that this is exactly the foreground binary star discussed above. Consequently, the linear radius of the nebula given in that catalogue, computed from the angular radius and the parallax of this star, has nothing to do with \ESO\ -- all the more so since, as noted in Section~\ref{txt:central_star}, this parallax itself does not differ significantly from zero.

The most reliable of the available estimates of the distance to \ESO\ have been obtained by statistical methods. \citet{2010ApJ...724..748H}, studying the oxygen abundance gradient in the Galactic disc, give for this PN (under the name M\,3$-$2) a heliocentric distance $D = 8.8$~kpc and a galactocentric distance $R_{\rm g} = 14.9$~kpc; close values ($D = 8.9$ and $R_{\rm g} = 15.0$~kpc) follow from the distance scale of \citet{2008ApJ...689..194S}. It is precisely this estimate that was used in \citet{2012AstL...38..707K} as one of the arguments in favour of a possible membership of \ESO\ in the CMa region. The accuracy of such statistical scales is not high and is usually estimated at the level of $\sim$30\%, and for bipolar PNe, to which \ESO\ belongs, they are the least reliable \citep{2016MNRAS.455.1459F}. Nevertheless, all of them place \ESO\ significantly farther away than the Sun and agree with the distance to the stellar overdensity in the CMa region ($\sim$7.2~kpc) within their uncertainties.

An independent check is provided by the total extent of the nebula measured by us. As shown in Section~\ref{txt:results}, \ESO\ is detected in the lines [\ion{N}{ii}]~$\lambda$6584~\AA\ and H$\alpha$ out to $\approx$55$^{\prime\prime}$, which at $D = 8.8$~kpc corresponds to a total linear size of $\approx$2.4~pc (radius $\approx$1.2~pc). Such a size is large but not exceptional for evolved bipolar PNe: comparable and larger sizes are known, for example, for evolved bipolar PNe in the solar neighbourhood \citep{2006MNRAS.372.1081F}. We note that the bright central part of \ESO\
has a size of only about 8$^{\prime\prime}$, which corresponds to $\approx$0.34~pc and is a completely typical value. Thus, the measured extent does not contradict a distance of $\sim$9~kpc, but neither does it allow it to be refined, since the linear sizes of PNe themselves vary within wide limits. Below, in the estimates of the kinematic age, we adopt $D = 8.8$~kpc
\citep{2010ApJ...724..748H}, explicitly indicating the dependence of the results on this value.

\subsection{The bipolar structure of \ESO\ in velocities and intensities}
\label{txt:structure}

As already noted in Section~\ref{txt:results}, \ESO\ has a bipolar structure. Such a structure is the result of the evolution of a binary star in the common-envelope phase \citep[for example,][]{1997MNRAS.284...32P,2009A&A...505..249M,2014MNRAS.441.2799C} and usually consists of two main elements: (1) an ejection of matter in the form of an expanding ring (torus) in the orbital plane of the stars of the binary system \citep[for example,][]{2006IAUS..234..139M,2015MNRAS.454..219G}; and (2) an ejection of matter in the direction perpendicular to the orbital plane of the stars of the binary system. In the second case the ejection of matter is either seen in the form of two cones, which creates the easily recognizable ``hour-glass'' shape \citep{1997MNRAS.284...32P}, or these ejections shape the nebula in the form of a ``barrel'' \citep{1993ApJ...404L..25F}.

In the case of \ESO\ both elements are present: both (1) and (2). Looking at Figure~\ref{fig:image}, one can note that the ring is located approximately in the region $-6^{\prime\prime} \le r \le +2^{\prime\prime}$ along the slit. Since the observed velocities at the north-eastern edge of the nebula are larger, this means that this edge is receding from the observer and, correspondingly, that the south-western edge is directed towards the observer. When analysing the velocity curve it is necessary to take into account the low resolution of our spectral data, as well as the projection effect, i.e. the understanding that the radiation of all the matter of the nebula falls into the slit: for example, both the near and the far walls of the cones and the matter between them.

The most reasonable approach appears to be to estimate the inclination of the bipolar structure to the line of sight from an analysis of the shape of the central ring. The ring looks like an ellipse with an axial ratio of approximately $7\!:\!10$, and this corresponds to an angle $i$ of the inclination of the bipolar structure to the line of sight of about 45~degrees ($i=0$ if the central ring looks circular, and $i=90$ when the ring is seen edge-on). On the velocity curve (middle panel of Figure~\ref{fig:vel}) this corresponds, in the H$\alpha$ line, to velocities of $\approx$63~\kms\ and $\approx$85~\kms, or to an expansion velocity of the ring of $\sim$16~\kms\ after correction for the angle $i$. This velocity value agrees very well with the measured expansion velocities of $\sim$15~\kms\ of similar rings in the extended planetary nebulae A\,79 and He\,2-428 \citep{2001A&A...377.1042R}.

When estimating the age of the ring it is necessary to take into account that the slit is set along the major axis of the nebula, i.e. along the axis of the cones, and therefore crosses the ring along the minor axis of the apparent ellipse. Consequently, the half-size of the ring observed along the slit, $\approx$4$^{\prime\prime}$, is not the true radius but its projection, equal to $R\cos i$, whence the true angular radius of the ring is $R \approx 4^{\prime\prime}/\cos i \approx 5\farcs7$. Adopting a distance to \ESO\ of $D = 8.8$~kpc (Section~\ref{txt:distance}), at which this radius corresponds to a linear radius of $\approx$0.24~pc, the kinematic age of this ring can be estimated as $\approx$1.5$\times$$10^4$~yr. It should be noted that in the line [\ion{N}{ii}]~$\lambda$6584~\AA\ the expansion velocity turns out to be larger, $\approx$28~\kms, and then the kinematic age turns out to be smaller, $\approx$8.5$\times$10$^3$~yr. Such a difference is not unexpected in itself: it means that the [\ion{N}{ii}] emission comes predominantly from the faster outer layers of the ring, whereas the emission in the H$\alpha$ line reflects the behaviour of the bulk of the nebula. Both estimates are proportional to the adopted distance and therefore have an uncertainty of no less than $\sim$30\%, inherited from the statistical distance scale.

As can be seen in the middle panel of Figure~\ref{fig:vel}, the velocity within the ring varies practically linearly. After that the velocity curve flattens into a slight velocity gradient in the H$\alpha$ line. The velocity values increase slightly towards the edges of the slit, and this apparently reflects the expansion of the cones of ejected matter. This picture is seen out to distances of $r\sim-12^{\prime\prime}$ and $r\sim+10^{\prime\prime}$, respectively. At these distances from the centre of the nebula, jumps in the velocities are seen, coinciding with an abrupt increase of the line ratios [\ion{N}{ii}]~$\lambda$6584/H$\alpha$ and [\ion{S}{ii}]~$\lambda$6716+6731/H$\alpha$, shown in the lower panel of Figure~\ref{fig:vel}, and with a sharp increase of the intensity of all these lines (upper panel of the same figure). On the whole, this picture fits well into the interpretation that it is precisely at these distances from the centre that we begin to see not only the walls of the cones but also their open ends, while the increase of the line ratios and of their intensities corresponds to a shock wave present at the edges of the expansion of these cones into the surrounding matter. The increase of the [\ion{N}{ii}]/H$\alpha$ and [\ion{S}{ii}]/H$\alpha$ ratios is a classical indicator of shock excitation in the spectra of PNe and \ion{H}{ii} regions \citep{1991PASP..103..815P}, and similar shock structures at the ends of bipolar outflows have been reliably detected spectroscopically in other objects as well. The closest example is the PN IC\,4634, for which \citet{2008ApJ...683..272G} showed, from long-slit spectroscopy, that the morphological, kinematic and emission properties of the observed bow-shock-like structures are consistent with the interaction of a collimated outflow with the surrounding matter. An analogous stratification of the emission in the [\ion{O}{iii}], H$\alpha$ and [\ion{N}{ii}] lines was observed in the bipolar knots of Hu\,1$-$2 \citep{2015MNRAS.452.2445F}. A systematic spectroscopic analysis of low-ionization structures in PNe and of the mechanisms of their excitation is given in \citet{2016MNRAS.455..930A} and \citet{2003ApJ...597..975G}. On the whole, the effect of observing the open ends of the cones terminates at distances of $r\sim-22^{\prime\prime}$ and $r\sim+20^{\prime\prime}$ from the centre, but faint emitting matter is seen farther out as well, to distances of $r\sim-26^{\prime\prime}$ and $r\sim+24^{\prime\prime}$, which apparently reflects earlier ejections of matter \citep[similar structures are observed in many bipolar PNe; see, for example,][]{1995A&A...293..871C,1998AJ....116.1357S}. In particular, it is interesting that at the south-western end of the slit the intensity of the lines [\ion{N}{ii}]~$\lambda$6584 and H$\alpha$ clearly shows one more wave of increase, which is consistent with the assumption of an earlier ejection of matter.

Our data also allow the kinematic age to be estimated in the second direction, perpendicular to the ring -- along the cones of ejected matter. The velocities at the edges of the cones differ from the systemic velocity of \ESO\ ($\approx$75~\kms) by $\pm$15--20~\kms, and the apparent distance from the centre to the edges of the cones is $\approx$15$^{\prime\prime}$. Since the axis of the cones is inclined to the line of sight by the same angle $i$, both observed quantities are distorted by projection: the true distance is equal to $r/\sin i$ and the true velocity to $\Delta V/\cos i$, so that in the ratio that determines the age the corrections almost completely cancel out, leaving a factor $\cot i \approx 0.98$. Hence the true expansion velocity of the cones is $\approx$21--29~\kms\ at a true length of $\approx$0.9~pc, which gives a kinematic age of $\approx$(3--4)$\times$10$^4$~yr. Even taking into account that both estimates
are proportional to the adopted distance and therefore share a common systematic, their ratio does not depend on the distance: the polar outflow turns out to be 2--3 times older than the central ring.

This result is not something exceptional. Thus, for the PN IPHASX\,J194359.5+170901 (the ``Necklace''), \citet{2011MNRAS.410.1349C} found that the kinematic age of the polar knots exceeds the age of the equatorial ring by a factor of two, and interpreted this as an indication that the polar outflow was formed \emph{before} the common-envelope phase, as a result of mass transfer onto the secondary component. Our data for \ESO\ give an age ratio of 2.1--2.7, i.e. they show an effect of the same sign and of practically the same magnitude. We note, however, an essential difference: in the ``Necklace'' the polar outflow is fast ($\sim$100~\kms\ against 28~\kms\ for the ring), whereas in \ESO\ the velocities in both directions are comparable (21--29~\kms\ against $\sim$16~\kms). In terms of the sample of \citet{2008AJ....135.2199D}, which covers bipolar PNe with different degrees of collimation, \ESO\ falls into the region of weakly collimated objects: the equatorial velocity measured by us lies at the upper boundary of the range observed by those authors (3--16~\kms), while the polar one lies in the very lowest part of their range (18--100~\kms). Thus, the large age of the cones in \ESO\ is caused not by a high outflow velocity but by their extent, which is consistent with the general picture of an evolved bipolar PN.

The result obtained also has a methodological significance. When estimating the kinematic age of bipolar PNe, the assumption of a homologous expansion ($V \propto r$) is often adopted, under which the ratio $r/V$, and hence the age, is the same in all directions by construction, so that a single kinematic age is assigned to the nebula. Thus, \citet{2012ApJ...751..116V}, studying the bipolar PN NGC\,2818, measured strongly different polar and equatorial expansion velocities ($V_{\rm pol} = 105$~\kms\ and $V_{\rm eq} = 20$~\kms), but, having adopted a homologous expansion, gave for it
a single age estimate $\tau_{\rm k} \simeq 8400\pm3400$~yr. Our data show that for \ESO\ such an assumption does not hold: the ages measured independently in two mutually perpendicular directions differ by a factor of 2--3. Since this ratio depends neither on the adopted distance nor, owing to the almost complete cancellation of the projection corrections, on the accuracy of the determination of the inclination angle, this conclusion appears to be robust. It means that an estimate of the kinematic age of a bipolar PN from a single direction alone can give a biased result, and that a separate measurement of the ages of the ring and of the polar outflow carries additional information about the history of the ejection of matter.

It is useful to compare \ESO\ with the PN Hu\,1$-$2, studied in detail in \citet{2015MNRAS.452.2445F}, since these two objects are similar in a whole set of key properties. Both nebulae are bipolar type~I PNe with enhanced abundances of helium and nitrogen, and both show reduced abundances of O, Ne, S and Ar compared with the mean values for PNe of the Galactic disc and bulge: for Hu\,1$-$2 this was noted by those authors, and for \ESO\ it follows from Table~\ref{tab:Major_che}. In both objects a strong \ion{He}{ii}~$\lambda$4686~\AA\ line is observed, while in the bipolar structures there is a stratification of the emission in the [\ion{O}{iii}], H$\alpha$ and [\ion{N}{ii}] lines, which \citet{2015MNRAS.452.2445F} interpret as the result of the combined action of shock waves and hard radiation from the central star. It is exactly such a picture that we observe in the open ends of the cones of \ESO\ (lower panel of Figure~\ref{fig:vel}). It is essential that the mechanical energy and the luminosity of the bipolar knots of Hu\,1$-$2 turned out to be comparable with those observed in PNe that are known to contain a close binary central star that has passed through the common-envelope stage, which is also consistent with the interpretation of the structure of \ESO\ proposed above.

At the same time, the differences between these objects are no less instructive. The bipolar knots of Hu\,1$-$2 move with velocities of $>$340~\kms, whereas the expansion velocities of \ESO\ observed by us are an order of magnitude smaller ($\sim$16--28~\kms), i.e. in \ESO\ we do not see fast collimated outflows. The orientation is different as well: the major axis of Hu\,1$-$2 lies within 10~degrees of the plane of the sky, while for \ESO\ we estimate an inclination $i\approx45^{\circ}$. Finally, the N/O ratio in \ESO\ is substantially higher than in Hu\,1$-$2 (N/O\,$\approx$\,0.9): for the central region ($+3^{\prime\prime}$) we obtain N/O\,$\approx$\,4.7, while the weighted mean value over all apertures (Table~\ref{tab:Major_che}) is N/O\,$\approx$\,3.7. We note that this value is reliable, in spite of the large value ICF(N)\,$\approx$\,3.9. Since in the technique used ICF(N)\,=\,O/O$^+$, the N/O ratio reduces to the directly measured ratio N$^+$/O$^+$ and practically does not depend on the correction for the unobserved ionization stages of nitrogen: for the region $+3^{\prime\prime}$, N/O\,=\,4.68 against N$^+$/O$^+$\,=\,4.77. In addition, both ionic abundances, N$^+$ and O$^+$, are computed using the directly measured temperature $T_{\rm e}$(\ion{N}{ii}), which additionally increases the reliability of this result. Thus, \ESO\ appears to be a more evolved and kinematically quieter analogue of Hu\,1$-$2, enriched in nitrogen even more strongly.

\subsection{The bipolar structure of \ESO\ in other parameters}
\label{txt:struc_par}

The availability of good spectroscopy also makes it possible to construct a picture of the various physical parameters and chemical abundances and to ``superimpose'' it on the bipolar structure of the nebula described in the previous section. As Tables~\ref{tab:Major_par} and \ref{tab:Major_che} and Figure~\ref{fig:2D_par} show, in spite of the variations of the temperatures and densities, the chemical abundances in the body of \ESO\ can be considered practically constant. At the same time, the situation with the chemical abundances in the SW direction from the centre looks much more homogeneous than in the NE one, where, starting from a distance of $r\sim+7^{\prime\prime}$, rather strong variations of all the parameters begin. The only simple explanation that suggests itself is the projection effect, since in this direction we simultaneously begin to see both the near wall of the cone and its open end directed away from the observer. In the SW direction the observer begins to see the open end of the cone of ejected matter and, through it, the far wall.

Let us estimate the significance of the observed variations quantitatively. For each element the hypothesis of a constant abundance along the slit was tested with the statistic $\chi^2 = \sum_i (x_i - \bar{x})^2/\sigma_i^2$, where $x_i$ is the value of 12+log(X/H) in the $i$-th aperture, $\sigma_i$ is its error from Table~\ref{tab:Major_che}, $\bar{x}$ is the weighted mean of these values, and the number of degrees of freedom is $\nu = N-1$. The resulting values of the reduced $\chi^2$ are given in the penultimate row of Table~\ref{tab:Major_che} and are $\chi^2/\nu$~=~3.0 (O), 3.4 (N), 6.0 (Ne), 3.2 (S), 6.8 (Ar) and 4.1 (Cl), which correspond to probabilities of a chance excess of $p < 4\times10^{-4}$. Formally, the hypothesis of constancy is rejected for all six elements, and not only for those whose variations are noticeable by eye (S, Ar and Cl), but also for oxygen and nitrogen.

By itself, however, this does not yet mean that the chemical composition varies along the nebula. The deviations from the means turn out to be strongly correlated between the different elements: the Pearson correlation coefficient between the deviations (in dex) for all pairs of elements, except the pairs involving the least reliably measured chlorine, lies in the range $r$~=~0.6--0.9. In other words, in those apertures where the sulphur abundance is ``enhanced'', the abundances of oxygen, neon and argon are simultaneously ``enhanced'' as well. Such coherence is naturally explained not by a chemical inhomogeneity but by a systematic effect common to all elements: the uncertainty of the electron temperature, which shifts the abundances of all elements coherently and in the same direction, as well as the decrease of the signal-to-noise ratio towards the edges of the slit.

Therefore a test of the constancy of the abundance ratios relative to oxygen, log(X/O), in which the common shift caused by the error of $T_{\rm e}$ largely cancels out, is more informative. The errors of these ratios were estimated conservatively, as $\sigma^2$(X/O)~=~$\sigma^2$(X)~+~$\sigma^2$(O), i.e. completely neglecting the correlation of the errors. Even with such an overestimation of the errors, the values of $\chi^2/\nu$ drop to 1.9 (N/O), 2.3 (Ne/O), 1.4 (S/O), 1.9 (Ar/O) and 2.0 (Cl/O) (the last row of Table~\ref{tab:Major_che}), i.e. the main part of the scatter is indeed common to all elements.

The remaining excess of $\chi^2$ is determined almost entirely by the single aperture $0^{\prime\prime}$, into which the bright background star falls (Section~\ref{txt:central_star}) and in which the formal errors are the smallest (0.02~dex): it gives deviations of $-5.5\sigma$ for neon and $-5.6\sigma$ for argon. It is precisely for this aperture that the subtraction of the stellar continuum introduces an additional systematic effect. If it is excluded from consideration, the hypothesis of a constant abundance ratio along the slit is no longer rejected for any element: $\chi^2/\nu$~=~1.4 ($p$~=~0.12) for N/O, 0.6 ($p$~=~0.89) for Ne/O, 1.3 ($p$~=~0.18) for S/O, 1.1 ($p$~=~0.31) for Ar/O and 2.2 ($p$~=~0.05) for Cl/O. The upper limit on any additional ``intrinsic'' scatter that would have to be added to the errors in order to obtain $\chi^2/\nu = 1$ is then 0.07~dex for N/O, S/O and Ar/O, 0.00~dex for Ne/O and 0.17~dex for Cl/O.

Thus, the variations of the S, Ar and Cl abundances that appear noticeable at first sight in Table~\ref{tab:Major_che} and in Figure~\ref{fig:2D_par} are not statistically significant as variations of the chemical composition: they follow the variations of oxygen and reflect the uncertainty of the electron temperature and of the ionization correction factors rather than a real chemical inhomogeneity of \ESO. The chemical abundances in the body of the nebula are constant along the slit to an accuracy of no worse than $\sim$0.1~dex.

We note that the conclusion about the constancy of the abundances along the nebula agrees with the results of the methodologically similar study of \citet{2003ApJ...597..975G}, who, from long-slit spectra along the major axis of the PN NGC\,7009, determined the physical conditions, the excitation and the chemical composition of all its morphological components -- the central ring, the jets and the knots. Those authors also found that the electron temperature remains practically constant over the whole nebula and that there are no noticeable changes of the He, O, Ne and S abundances between its components. Thus, the picture observed in \ESO\ -- variations of the physical conditions with constant abundances -- is apparently typical of PNe with a complex morphology.

In the central part of the PN, $-8^{\prime\prime} \le r \le +4^{\prime\prime}$, the temperatures $T_{\rm e}$(\ion{O}{iii}) and $T_{\rm e}$(\ion{N}{ii}), measured using the faint auroral lines [\ion{O}{iii}]~$\lambda$4363 \AA\ and [\ion{N}{ii}]~$\lambda$5755 \AA, can be considered constant and equal to $17460\pm230$~K and $13650\pm340$~K, respectively (here and below the weighted mean values over the apertures of this region are given). In this region $T_{\rm e}$(\ion{O}{iii}) is significantly larger than $T_{\rm e}$(\ion{N}{ii}), which is the usual picture for the two-zone structure of a PN. Farther out the ionization level weakens and at the distances $-9^{\prime\prime}$ and $+5^{\prime\prime}$ the temperatures become equal, and beyond that they can be considered equal to each other within the errors. The electron density along the slit remains practically constant and amounts to $\approx$250--380~cm$^{-3}$. It possibly increases at the open ends of the cones, but these values have rather large errors.

The abundances obtained make it possible to return to the results of \citet{1998A&A...332..721P}, in which the chemical composition of a sample of 15 bipolar PNe was studied and in which \ESO\ was included under the name M\,3$-$2. In order to avoid the influence of the corrections for unobserved ionization stages, those authors used the directly measured ratio N$^+$/O$^+$, constructing for their sample the diagram log(N$^+$/O$^+$) -- log(O/H) (their Fig.~9). For M\,3$-$2 they were able to measure only one point in the centre of the nebula, and the obtained value
log(N$^+$/O$^+$)~=~0.5$\pm$0.2 had a large uncertainty. Our data give for the region $-3^{\prime\prime}$ log(N$^+$/O$^+$)~=~0.61$\pm$0.08, i.e. they confirm their result, but with a 2.5 times better accuracy, and, in addition, they make it possible to trace this ratio along the whole major axis of the nebula. It is essential that \ESO\ fits well into the trend noted by \citet{1998A&A...332..721P}: at log(N$^+$/O$^+$)$>$0.2 the bipolar PNe of their sample have reduced oxygen abundances, which is exactly what we observe (12+log(O/H)~=~7.84$\pm$0.02).

Helium deserves a separate mention. \citet{1998A&A...332..721P} noted that M\,3$-$2 is one of the three most helium-rich PNe of their sample (together with He\,2$-$111 and NGC\,6537), and that the theoretical models of that time did not reproduce such large helium excesses. The value He/H~=~0.229$\pm$0.007 obtained by us (Table~\ref{t:Chem1}) independently confirms this
conclusion. Moreover, those authors drew attention to the fact that in bipolar PNe the N/O ratio grows only up to a value of log(N/O)$\sim$0.5, after which a further enrichment manifests itself only in the growth of He/H. The values found by us, log(N/O)~=~0.57$\pm$0.03 (weighted mean over all apertures) and log(N$^+$/O$^+$)~=~0.61, together with the high helium abundance, place \ESO\ exactly in this region of the diagram, i.e. among the bipolar PNe with the most massive progenitors.

The Galactic foreground extinction in the direction of \ESO\ is $E(B-V) = 0.23$~mag \citep{2011ApJ...737..103S}. A comparison of this value with the one obtained from our spectroscopy (lower panel of Figure~\ref{fig:2D_par}) shows that in the very central part of the PN it is close to the foreground value and grows towards the edge, reaching maxima of $E(B-V) \sim 0.4$~mag in the region of the open ends of the cones.

\section{Conclusions}
\label{txt:summ}

This paper presents the results of detailed long-slit spectroscopy of the bipolar PN \ESO, obtained with the RSS spectrograph of the SALT telescope in the spectral range 3500--7400~\AA\ with the slit positioned along the major axis of the nebula. The main results can be formulated as follows.

\begin{enumerate}

\item For the first time the physical conditions and the chemical abundances have been determined not for the nebula as a whole, but separately for 19 regions along its major axis over an extent of 40$^{\prime\prime}$, i.e. for the different morphological parts of \ESO: the central ring,
the cones of the bipolar outflow and their open ends. The use of the $T_{\rm e}$--method with a direct measurement of both auroral lines, [\ion{O}{iii}]~$\lambda$4363~\AA\ and [\ion{N}{ii}]~$\lambda$5755~\AA, made it possible to compute two independent electron temperatures in all these regions.

\item The kinematics along the slit directly confirms the bipolar structure of \ESO. For the first time the inclination of the nebula to the line of sight has been estimated as $i \approx 45^{\circ}$, as well as the projection-corrected expansion velocity of the ring in the H$\alpha$ line as $\sim$16~\kms. At the adopted distance $D = 8.8$~kpc this corresponds to a kinematic age of the ring of $\approx$1.5$\times$10$^4$~yr. The kinematic age has also been estimated in the second direction, perpendicular to the ring, as $\approx$(3--4)$\times$10$^4$~yr, i.e. the polar outflow is 2--3 times older than the central ring. The ratio of these ages does not depend on the adopted distance. The weighted mean heliocentric velocity of \ESO\ is V$_{\rm hel}$=74.7$\pm$1.6~\kms, and the measured velocity depends substantially on which part of the nebula falls into the slit.

\item At distances of $r\sim-12^{\prime\prime}$ and $r\sim+10^{\prime\prime}$ from the centre, jumps of the velocity are observed, coinciding with an abrupt growth of the ratios [\ion{N}{ii}]~$\lambda$6584/H$\alpha$ and [\ion{S}{ii}]~$\lambda$6716+6731/H$\alpha$ and with a sharp increase of the intensities of these lines. This is naturally interpreted as the transition from observing the walls of the cones to observing their open ends, while the growth of the line ratios is interpreted as a manifestation of a shock wave at the boundary of the expansion of the cones into the surrounding matter.

\item In spite of the variations of the temperatures and densities, which show the stratification of the nebula, the chemical abundances in the body of \ESO\ can be considered practically constant along the slit. A test with the $\chi^2$ statistic shows that the scatter of the absolute abundances ($\chi^2/\nu$~=~3--7) is common to all elements and is caused by the uncertainty of the electron temperature: the deviations from the means are correlated between the elements with $r$~=~0.6--0.9, while for the abundance ratios relative to oxygen the hypothesis of constancy is not rejected for any element ($\chi^2/\nu$~=~0.6--1.4, $p$~=~0.12--0.89). The upper limit on any real inhomogeneity of the chemical composition is $\sim$0.1~dex.

\item An independent analysis of the spectrum of the star projected close to the centre of \ESO\ gives $T_{\rm eff}$ and $\log g$ that agree
with the results of \citet{2018A&A...619A..84B}. At the same time, the colour excess $E(B-V)$ and the heliocentric velocity of this star differ sharply from the values obtained for the nebula itself. Together with its non-central position, this independently confirms that this star is a nearby foreground star and is not the central star of \ESO.

\end{enumerate}

\section*{ACKNOWLEDGEMENTS}

All observations presented in this paper were obtained with the Southern African Large Telescope (SALT), under the observational programme 2012-2-RSA-001 (PI: Kniazev). The author thanks the National Research Foundation of South Africa (NRF) for the support of this work.

\bibliographystyle{aspb1}
\bibliography{SALT_ESO428.bib}

\end{document}